\documentclass[aps,prl,twocolumn,superscriptaddress,nofootinbib]{revtex4-2}
\usepackage[english]{babel}
\usepackage[colorlinks,urlcolor={blue}]{hyperref}
\usepackage{eurosym}
\usepackage[usenames]{color}
\usepackage{graphicx}
\usepackage{amsmath}
\usepackage{amsfonts}
\usepackage{amssymb}
\usepackage{mathrsfs}
\usepackage{bm}
\usepackage{verbatim}
\usepackage{ulem}
\usepackage{enumitem}   

\usepackage{dcolumn}
\usepackage{subfigure}

\usepackage{multirow}

\begin{document}

\title{\textbf{Calculations for nuclear matter and finite nuclei within and beyond energy--density--functional
theories through interactions guided by effective field theory}}
\author{C.-J. Yang}
\affiliation{Department of Physics, Chalmers University of Technology, SE-412 96, G{\"o}%
teborg, Sweden}
\affiliation{Nuclear Physics Institute of the Czech Academy of Sciences, 
              25069 \v{R}e\v{z}, Czech Republic}
\author{W.G. Jiang}
\affiliation{Department of Physics, Chalmers University of Technology, SE-412 96, G{\"o}%
teborg, Sweden}
\author{S. Burrello}
%\affiliation{Universit\'e Paris-Saclay, CNRS/IN2P3, IJCLab, 91405 Orsay, France}
\affiliation{Institut f\"ur Kernphysik, Technische Universit\"at Darmstadt, 64289 Darmstadt, Germany}
\author{M. Grasso}
\affiliation{Universit\'e Paris-Saclay, CNRS/IN2P3, IJCLab, 91405 Orsay, France}

\begin{abstract}
We propose a novel idea to construct an effective interaction under energy-density-functional (EDF) theories 
which is adaptive to the enlargement of the model space. Guided by effective field theory principles, iterations 
of interactions as well as enlargements of the model space through particle-hole excitations are carried 
out for infinite nuclear matter and selected closed-shell nuclei ($^4$He, $^{16}$O, $^{40}$Ca, $^{56}$Ni and $^{100}$Sn) up to next-to-leading order. 
Our approach provides a new way for handling the nuclear matter and finite nuclei within the same scheme, with advantages from both EDF and \textit{ab initio} approaches.

\end{abstract}

\maketitle

\affiliation{Department of Physics, Chalmers University of Technology,
SE-412 96, G{\"o}teborg, Sweden}

\affiliation{Department of Physics, Chalmers University of Technology,
SE-412 96, G{\"o}teborg, Sweden}

\affiliation{Universit\'e Paris-Saclay, CNRS/IN2P3, IJCLab, 91405 Orsay,
France}

\affiliation{Universit\'e Paris-Saclay, CNRS/IN2P3, IJCLab, 91405 Orsay,
France}

\bigskip
\section{Introduction} 
One important challenge in the nuclear many-body problem concerns the construction of interactions. Existing state-of-the-art approaches can be mainly categorized into two extremes: one starts with bare nucleon-nucleon (NN) degrees of freedom and improves the results order by order following effective-field-theory (EFT) \cite%
{We90,We91,bira,bira1,Epel,Epel1,idaho,idaho1,Epelmore,Epelmore2,reviews,reviews2,reviews3,rev1} through \textit{ab initio} calculations \cite%
{dickhoff2004,lee2009,bogner2010,barrett2013,Carbone:2013eqa,hagen2014,HERGERT2016165, RevModPhys.87.1067, BARNEA1999427, GLOCKLE1996107}; the other adopts the energy-density-functional (EDF) framework to build an ``in medium" interaction within the self-consistent mean-field (MF) approximation. However, both approaches suffer from longstanding shortcomings.

Indeed, although \textit{ab initio} approaches allow to construct the interaction on a clear foundation, they still suffer from technical difficulties concerning the reduction of the enormous model space required to converge the many-body calculations\cite{az,az2,az3,az4,az5,az6,az7,az8,az9,trap,srg,srg2,srg3}. Attempts along this direction have been carried out through methods of unitary transformations \cite{srg,srg2,srg3} or an EFT procedure which accounts for both ultraviolet and infrared truncations \cite%
{az,az2,az3,az4,az5,az6,az7,az8,az9}. Moreover, much effort has been spent to face theoretical problems related to the power counting issues \cite{bira_problem,de1,Griesshammer:2021zzz,vanKolck:2021rqu} and the growing importance of three- and four-nucleon forces with the number of particles in the system \cite{yangproblem,yang20,Yang:2021vxa}. Nonetheless, a definite solution to these questions is far from being assured.

On the other hand, a strong model-dependence characterizes the effective interactions usually employed in the EDF framework, as derived at the MF level. To complicate matters, it is known that beyond MF (BMF) effects need also to be taken into account. In contrast to the non-perturbative treatment adopted in \textit{ab initio} calculations, approaches
such as the MF Hartree-Fock approximation or BMF methods are then applied (see for instance Refs. \cite%
{gamba2015,gamba2020,bmf,bmf2,bmf3,bmf4,pvc,yglonuclei,burrello2019,burrello2019fro}). However, BMF effects are usually evaluated by employing the same interaction fitted at MF, which generates an overcounting of correlations at the BMF level. More refined methods exist to overcome this problem, such as self-energy-subtraction procedures, which are used for example in the second random-phase approximation \cite%
{gamba2015}.
Nevertheless, there is a lack of an order-by-order organization scheme to generate effective interactions applicable to both nuclear matter and finite nuclei. Even more importantly, as a common drawback of both EDF and \textit{ab initio} approaches, the interaction is defined in a fixed model space---which stays \textit{unchanged} throughout all considered orders.

Inspired by recent efforts toward bridging EDF and EFT ideas \cite%
{hammer,Kaiser:2015vpa,Kaiser:2017xie,mog,Moghrabi:2016hnp,yang2017-a,
yang2017-b,yang2016,orsay5,orsay6,orsay7,orsay8,orsay9,orsay10,kid,kid2,kid3,kid4,kid5,dick,Burrello2020,marini,DeGregorio:2022anr,PhysRevC.75.044312,PhysRevC.78.054308,PhysRevC.85.014313,PhysRevC.93.044314,PhysRevC.94.061301,PhysRevC.95.034327,De_Gregorio_2017,PhysRevC.97.034311,PhysRevC.99.014316,PhysRevC.101.024308,DeGregorio:2021dsr}, 
we probe in this work a novel possibility---which proposes to improve \textit{both} the interaction and the model space order by order.
Specifically, we assume there is an underlying EFT expansion where the MF results correspond to the LO contribution.
Subleading corrections are then added, which contain the iterated LO interaction renormalized in an enlarged model space through particle-hole excitations. 
Constructing an EFT in this direction naturally leads to a novel setup
which demands: 
\begin{itemize}
\item The interaction to be adaptive to
the growth of the model space at each order. 
\item Iterations of LO interactions to be performed through an in-medium
propagator.
\end{itemize} 
This strategy was already applied to infinite matter for instance in Refs. \cite{mog,yang2016,orsay7,Burrello2020}.
Note that an attempt to include the
second-order Dyson diagrams have been
proposed and applied to the calculation of the $^{16}$O binding energy
in Ref. \cite{Brenna2014}. 
However, an investigation that fully exploits the advantages
of an enlarged model space and analyzes the renormalizability of various power-counting scenarios
for both nuclear matter and finite nuclei is so far absent.

We present here a first study where we apply such a strategy to both matter and finite nuclei, putting the basis for a novel approach to be adopted in nuclear structure calculations.
Our focus is indeed to develop a unified framework together with an
order-by-order improvable and renormalizable interaction which has the potential to be applied
to infinite nuclear matter and nuclei across the entire nuclear chart, as traditional EDF does.

\section{Leading order: Empirical interactions renormalized under mean field model space}
We start by defining the Hamiltonian $H_{LO}$, which contains the kinetic
term plus the LO interaction term $V^{LO}$ 
\begin{equation}
H_{LO}=\sum_{i}e_{i}\widehat{n}_{i}+\sum_{i>j}V_{ij}^{LO},  \label{h}
\end{equation}%
where $e_{i}$ and $\widehat{n}_{i}$ are the energy and the particle-number
operator for the particle $i$. The interaction term $V_{ij}^{LO}$ is a two-body
operator to be determined. To speculate a reasonable LO interaction under
EDF, we make use of one basic requirement of EFT---the renormalizability of
the observables. Studies performed for nuclear matter in Refs. \cite%
{yang2017-a,yang2017-b,Burrello2020} suggest that a $t_{0}-t_{3}$ model of
Skyrme-type interactions is most likely to be a suitable candidate for $V^{LO}
$. MF calculations of Eq. (\ref{h}) are straightforward for both nuclear
matter and finite nuclei.

However, we do not adopt the conventional Hartree-Fock procedure here. 
Guided by empirical information (such as the information obtained by
shell-model calculations fitted to experiments), one could start with an
ansatz of the wavefunction $\Psi $ and evaluate $H_{LO}$ by calculating its
matrix element. Note that here $\Psi $ defines our model space at LO and
does not change with the effective interaction. Ref. \cite%
{Jiang2018} showed that reasonably good results can be obtained by directly
evaluating the Gogny interaction between $\Psi $ consisting of a single-particle basis constructed in the shell model. Inspired by that, we directly
define our LO model space as the shell-model wavefunction up to the highest
occupied shell and calculate the expectation value of the Hamiltonian
perturbatively. The ground-state (g.s.) energy of the system at LO can be
written as $E_{g.s.}^{LO}=E_{v}+E_{coul}-t_{CM}+E_{c}$, with $%
E_{v},E_{coul},E_{c}$ the energy of valence particles, the Coulomb, and the core contributions,
respectively; $t_{CM}=\frac{3}{4}\hbar \omega $ is the center-of-mass (CM)
kinetic energy. Throughout this exploratory work we only consider closed-shell
nuclei, so that $E_{v}=0$, and Coulomb is treated within mean-field approximation. The core energy can be further written as the core
kinetic plus the core potential energy, that is $E_{c}=t_{c}+V_{c}$, where \cite{Jiang2018}%
\begin{eqnarray}
V_{c} &=&\sum_{j_{a}^{c}\leq j_{b}^{c}}\sum_{JT}(2T+1)(2J+1)\left\langle
j_{a}^{c}j_{b}^{c}JT|V^{LO}|j_{a}^{c}j_{b}^{c}JT\right\rangle .  \notag \\
t_{c} &=&\sum_{j_{a}^{c}}(2T+1)(2J+1)\left\langle j_{a}^{c}|\widehat{t}%
|j_{a}^{c}\right\rangle ,  
 \label{core}
\end{eqnarray}%
Here $j_{a}^{c},j_{b}^{c}$ label the single-particle orbits in the core, $%
\widehat{t}$ is the kinetic energy operator; $J$ and $T$ are the total angular
momentum and isospin quantum number for each pair of interacting
particles, respectively. 

%Since the interaction is constructed in relative coordinates, one needs Moshinsky transformation\cite{moshinsky,moshinsky2} to transform the
%single particle, Harmonic oscillator (HO) basis into its relative and
%center-of-mass (CM) components.
%\begin{widetext}
%\begin{eqnarray}
%\mid (n_{a}l_{a}m_{a})(n_{b}l_{b}m_{b})\lambda \mu \rangle
%=\sum\limits_{nlNL}M_{\lambda }(nlNL;n_{a}l_{a}n_{b}l_{b})\mid
%(nlm)(NLM)\lambda \mu \rangle ,  \label{mosh}
%\end{eqnarray}
%\end{widetext}
%where $\lambda $ is the total orbital angular momentum and $%
%\mu $ is its z-component. $M_{\lambda }(nlNL;n_{a}l_{a}n_{b}l_{b})$ is given
%by the Moshinsky bracket\cite{moshinsky}. $nlm$ ($NLM$) are the principle,
%angular momentum, and z-component of the angular momentum quantum number in
%the relative (CM) coordinates. 
Note that the combination of the harmonic oscillator (HO) strength $%
\hbar \omega $ and $N_{max}$ (denoting the truncation up to the
highest occupied shell) provides a natural cutoff of the Fermi sphere in
finite nuclei and might play a similar role as the Fermi momentum $k_F$ in the nuclear matter case. Since
our interaction is singular, without additional regulators, results in general will not converge with the
increase of $\hbar \omega $. For each nucleus, there exists an optimal $\hbar
\omega $ so that the shell-model basis matches the size of the nucleus. For
nuclei with mass number $A$, the empirical $\hbar \omega \approx 45A^{-1/3}-25A^{-2/3}$ is frequently adopted \cite{ Blomqvist1968}. 
 With the above equations, evaluations of the g.s. the energy of $^{4}$He, $^{16}$O and $^{40}$Ca using $V^{LO}$ are
straightforward. The detailed derivation is given in Refs. \cite%
{Jiang2018,Brenna2014} and summarized in the supplemental materials, together with the form assumed by the adopted LO interaction. 
\begin{figure}[tbp]
\includegraphics[scale=0.32]{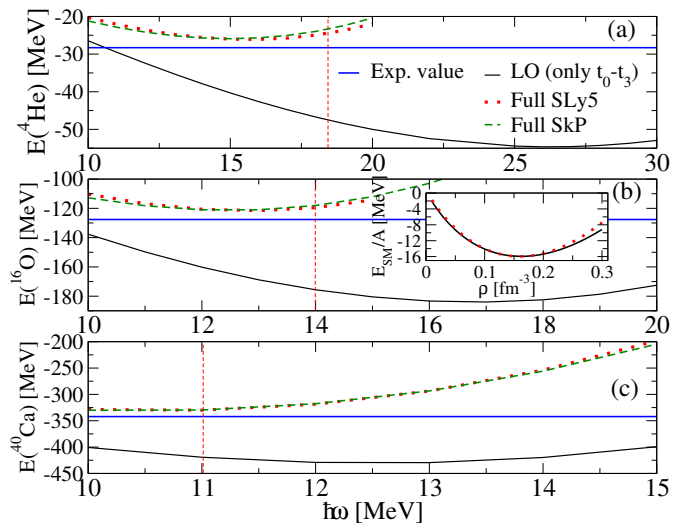}
\caption{Ground state energies of $^{4}$He (a), $^{16}$O (b), and $^{40}$%
Ca (c) as a function of $\hbar \protect\omega $. Results obtained from a $t_0$-$t_3$ model and full SLy5,
and SkP functionals are plotted as black solid, red dotted, and green dashed lines, respectively. The
empirical $\hbar \protect\omega $ value for each nucleus is marked as a red
vertical dashed line. 
The horizontal blue lines represent the experimental energies.
The SM energy per particle at LO is plotted as a function of
the density $\protect\rho $ in the inset.}
\label{figLO}
\end{figure}

We
present the g.s. energies as a function of $\hbar \omega $ in Fig.\ref{figLO},
where a $t_0-t_3$ model of the SkP parametrization \cite{skp} is adopted for $V^{LO}$. 
LO calculations systematically provide strongly overbound nuclei
with respect to experimental data, even at the empirical value of $\hbar
\omega $, though the corresponding MF equation of state (EoS) for symmetric matter (SM) (shown in the inset of Fig.\ref{figLO}) 
is quite satisfactory.
This is not surprising judging from the simple form of the LO interaction.
We have tried other $t_{0}-t_{3}$ parametrizations, which reproduce as well the
empirical SM EoS, and found that the systematic
overbinding persists.
 On the other hand, with the $t_{1,2}$ Skyrme terms
included, the MF g.s. energies obtained from SkP \cite{skp} or SLy5 \cite{cha1,cha3} are very reasonable, 
with the minimum also located close to the empirical $\hbar \omega $ value.
Although an EFT should aim to capture the most important physics already
at LO, one could argue
that basic physics is roughly captured once the equation of state of symmetric matter can be reproduced up to
saturation density.

\section{Next-to-leading order: Two possibilities of improvements}

\begin{figure}[tbp]
\includegraphics[scale=0.15]{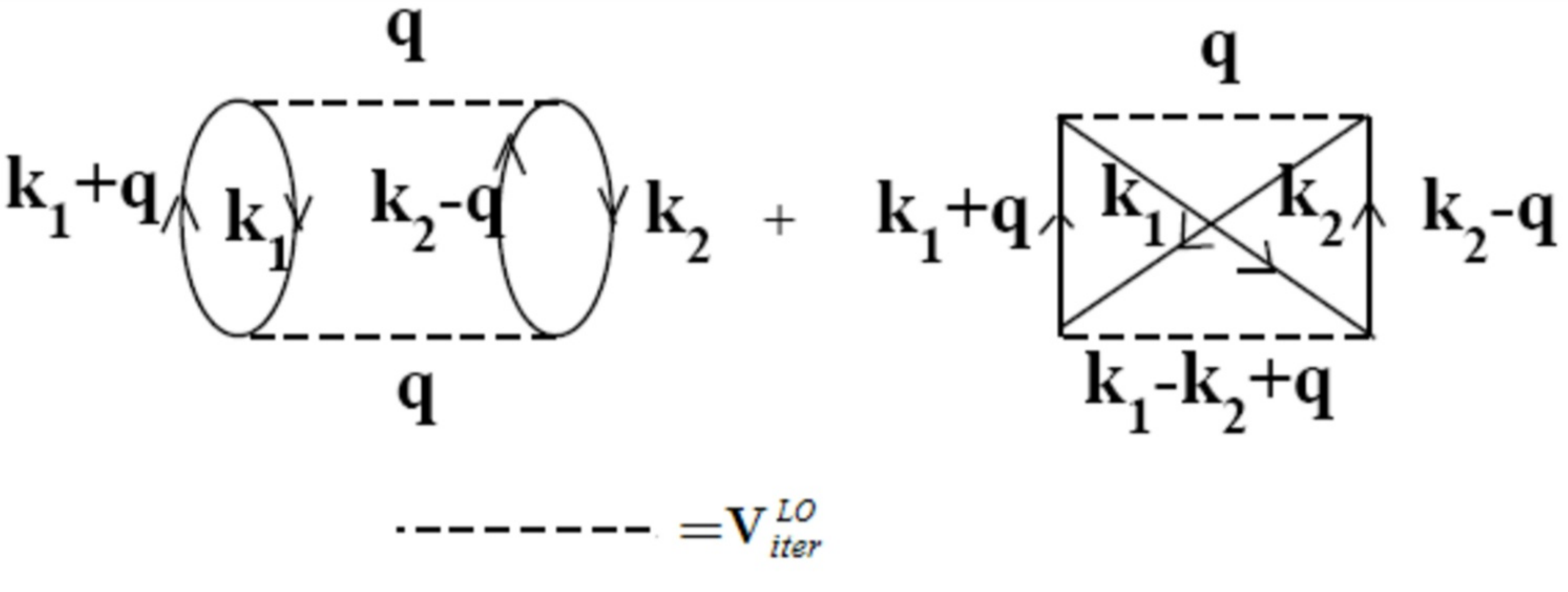}
\caption{Once-iterated diagrams for the interaction $V_{iter}^{LO}$. $\mathbf{k}_{1}$ ($%
\mathbf{k}_{2}$) denotes the single-particle momentum of the initial (final)
state, and $\mathbf{q}$  is the transferred momentum.}
\label{NLOdiagram}
\end{figure}

To improve further, two approaches are possible. First, one could add more terms to the effective interaction together with an EFT-based
speculation on their form or importance, and then evaluate again Eq. (\ref{h}) to produce
better fits to a wider range of nuclear properties. Many attempts have been devoted toward this direction\cite{kid,kid2,kid3,kid4,kid5,PhysRevC.96.044330,2017,PhysRevC.97.044304}.
On the other hand, since nuclei are bound-states, they should be generated by at least partial iteration of a certain interaction. The MF description might be then improved by considering higher-order corrections coming from the iterated diagrams, such as the ones corresponding to the enlargement of the model space through particle-hole excitations.
Improvements in this direction are much less studied, as iterating effective interactions built at MF-level in the loops usually generates self-consistency problems, unless
renormalization is taken care properly.

Starting from next-to-leading order (NLO), we improve our theory by considering both of the above directions, and demonstrate how to establish 
self-consistent NLO corrections through proper renormalization procedures.
Up to NLO, one has 
\begin{equation}
E_{NLO}=E^{LO}+\left\langle \Psi \left\vert V_{ij}^{CT}\right\vert \Psi
\right\rangle +E_{iter}^{NLO},  \label{nlo1}
\end{equation}%
where $V_{ij}^{CT}$ is the higher-order contact interaction entering at NLO
with its contribution evaluated at the MF level (the same way as in Eq. (\ref{core}%
)). The structure of $V_{ij}^{CT}$ has to be determined according to the
renormalizability and the power-counting scheme. $E_{iter}^{NLO}$ represents
the contribution of the once-iterated diagrams listed in Fig.\ref{NLOdiagram}%
. The general form of $E_{iter}^{NLO}$ reads \cite{baranger}:%
\begin{equation}
E_{iter}^{NLO}=-\frac{1}{4}\sum_{j_{a}^{c}\leq j_{b}^{c},X_{\alpha }\leq
X_{\beta }}\frac{\left\vert \left\langle
j_{a}^{c}j_{b}^{c}JT|V_{iter}^{LO}|X_{\alpha }X_{\beta }JT\right\rangle
\right\vert ^{2}}{\varepsilon _{\alpha }+\varepsilon _{\beta }-\varepsilon
_{a}-\varepsilon _{b}},  \label{nlo2}
\end{equation}%
where $j_{a(b)}^{c}$ are the same as in Eq. (\ref{core}) because one 
stops at the highest occupied orbital; $X_{\alpha (\beta )}$ stands for excited
states, where the summation starts at the Fermi sphere and stops at an upper
limit which defines the second-order model space; $\varepsilon
_{i}=k_{i}^{2}/2m$ is the single-particle energy of each state having momentum $k_i$ (the
effective mass is set to its bare value $m =$ 939 MeV in this work). $V_{iter}^{LO}$
denotes the part of the LO interaction which is iterated to provide the NLO contribution. A straightforward evaluation of Eq. (\ref{nlo2})
is in principle possible. However, the truncation applied to the excited
states in the single-particle basis cannot be directly matched with the truncation
performed for the EoS of matter in Refs. \cite{yang2017-a,yang2017-b,Burrello2020}%
---where a relative momentum cutoff $\Lambda $ is applied.
Moreover, Moshinsky transformations require all excited states $%
X_{\alpha ,\beta }$ to be represented in terms of the HO basis\footnote{%
Alternatively, one could go through extra processes which involve
the decomposition of the chosen basis into the HO one \cite{gammashell,gammashell2}.%
}, which complicates the matching between different nuclei, as they
correspond to different $\hbar \omega $ and $X_{\alpha (\beta )}$%
. 

To produce a renormalized interaction to be easily applied to all cases, we
proceed as follows. First, since excitations are governed by $V_{iter}^{LO}$%
, one can directly represent the relevant wavefunctions in relative
coordinates. Let us call $\mathbf{k}_{1}$, $\mathbf{k}_{2}$ ($\mathbf{k}%
_{1}^{\prime }$, $\mathbf{k}_{2}^{\prime }$) the single-particle momenta of
the initial/final (intermediate) state. Then, the incoming and outgoing
momenta in relative coordinates are $\mathbf{k}=(\mathbf{k}_{1}-$ $\mathbf{k}%
_{2})/2,$ $\mathbf{k}^{\prime }=(\mathbf{k}_{1}^{\prime }-$ $\mathbf{k}%
_{2}^{\prime })/2+\mathbf{q}$, where $\mathbf{q}$ is the transferred
momentum. Eq. (\ref{nlo2})
can be rewritten as%
\begin{widetext}
\begin{eqnarray}
E_{iter}^{NLO}=\frac{f(\hbar \protect\omega)}{4}\left[ \int d^{3}\mathbf{k}%
\int d^{3}\mathbf{k}^{\prime }\int d^{3}\mathbf{K}%
\left\langle \Psi (\mathbf{K};\mathbf{k})\left\vert V_{iter}^{LO}(\mathbf{k};\mathbf{k}^{\prime})\right\vert
\psi (\mathbf{k}^{\prime })\right\rangle G\left\langle \psi (\mathbf{k}%
^{\prime })\left\vert V_{iter}^{LO}(\mathbf{k}^{\prime};\mathbf{k})\right\vert \Psi (\mathbf{K};\mathbf{k}%
)\right\rangle \right] _{BC}.  \label{nlo3}
\end{eqnarray}
\end{widetext}where 
\begin{equation}
G=\frac{-m}{k^{\prime 2}-k^{2}}.  \label{propagator}
\end{equation}%
$\Psi $ is represented in the same basis used at LO, which depends on the CM
momentum $\mathbf{K} = \mathbf{k}_1 + \mathbf{k}_2 = \mathbf{k}_1^{\prime} + \mathbf{k}_2^{\prime}$ and on the relative momentum $\mathbf{k}$. $\psi
=\sum_{i}\phi _{i}$ denotes the intermediate excitations, where $\phi _{i}$
can be represented by any complete basis. One caveat is that if one chooses
to expand $\Psi $ and $\psi $ in a different basis, an overall factor $f\neq 1$
will be needed to fix the norm. To define the
intermediate model space, one must truncate it either by the number of basis states or by 
the highest momentum. In this work, we choose the second option and adopt the free
wave-packets basis so that % $\psi (k^{\prime })=\frac{1}{(2\pi )^{3}}%
%\delta (k^{\prime }-k)$. In this way, 
$f$ depends only on $\hbar \omega $.
The detailed derivation leading to Eq. (\ref{nlo3}) is given in the supplemental materials. 
Note that the conversion of initial/final and intermediate single-particle-basis states (which are also restricted as mentioned before) to relative
coordinates results in a boundary condition (BC) which couples $k_{F}$ to new variables $\mathbf{k}$, $\mathbf{k}^{\prime }$ and $%
\mathbf{K}$. The 3-folded integral under the same BC has been carried out
to obtain the second-order EoS \cite{baranger,yang2016}, and is to be carried
out in a similar manner in Eq. (\ref{nlo3}). However, unlike the nuclear
matter case---where a clear definition of Fermi
momentum is possible---$k_{F}$ is not clearly given in finite nuclei. In the nuclear matter case,
the radial integral $dk$ is truncated by $k\in[0,k_{F}]$. 
On the other side, in finite nuclei, the same integrals are carried out through $k\in[0,\infty]$. However, the shell structure (the LO wavefunctions of a nucleus at a chosen $\hbar\omega$) provides 
a natural truncation analogous to $k_F$. To
proceed, we interpret $k_{F}$
in finite nuclei to be the highest momentum each wavefunction can
access. The procedure to extract $k_{F}$ in a finite nucleus is thus the
following. First, we evaluate the g.s. energy at the MF level and separate the
contributions from the $t_{0}$, the $t_{3}$, and the $V^{CT}$ terms for each
nucleus. Then, we compare the ratios $\frac{\left\langle
t_{0}\right\rangle }{\left\langle t_{3}\right\rangle }$ and $\frac{%
\left\langle t_{0}\right\rangle }{\left\langle V^{CT}\right\rangle }$ to
their corresponding values in SM. %\footnote{%
%All three nuclei we considered ($^{4}$He, $^{16}$O and $^{40}$Ca) have the
%same number of protons and neutrons. Therefore, SM should be adopted in the
%comparison.}. 
The ratios in nuclear matter depend on $k_{F}$, whereas the same ratios in finite nuclei are related to their shell structure. By requiring the same ratios between finite nuclei and nuclear matter, we
can extract the corresponding $k_{F}$ for $^{4}$He, $^{16}$O and $^{40}$%
Ca (denoted as $k_A$) under various $\hbar \omega $ values. The resulting $k_{A}$ are listed in Table \ref{kf_list}. 
\begin{table}[h]
\begin{tabular}{ccccccccc}
\hline\hline
$\hbar\omega $ (MeV) & 11 & 12 & 13 & 14 & 15 & 16 & 17 & 18 \\ \hline
$k_{A}$ of $^{4}$He & 0.90 & 0.95 & 0.98 & 1.02 & 1.05 & 1.08 & \textbf{1.12} & 1.15 \\ 
$k_{A}$ of $^{16}$O & 1.08 & 1.13 & 1.18 & \textbf{1.22} & 1.26 & 1.30 & 1.34 & 1.38 \\ 
$k_{A}$ of $^{40}$Ca & \textbf{1.25} & 1.37 & 1.35 & 1.40 & 1.45 & 1.49 & 1.54 & 1.58 \\ \hline\hline
\end{tabular}%
\caption{$k_{A}$ (unit: fm$^{-1}$) for $^{4}$He, $^{16}$O and $^{40}$Ca under various $\hbar 
\protect\omega$. Those adopted in Fig. \ref{figNLO} are highlighted by bold text. }
\label{kf_list}
\end{table}
One can see that a heavier nucleus (and a larger $\hbar \omega$) naturally corresponds to a higher $k_A$. 
We have tried several interactions having different values of $\alpha$ (the power of the density in the $t_3$ term) and found a very weak spreading 
($\le$ 1 \% variations for $\alpha\approx 0.16-0.3$)\footnote{For $\alpha$ up to 1, the 
extracted $k_F$ can vary up to $ 5\%$. } between the values of $k_A$ obtained by using such interactions in the matching of the ratios.

%This confirms that $k_{F}$
%extracted in this way indeed corresponds to the highest momentum mode in the
%finite nuclei wavefunction.
\begin{figure}[tbp]
\includegraphics[scale=0.32]{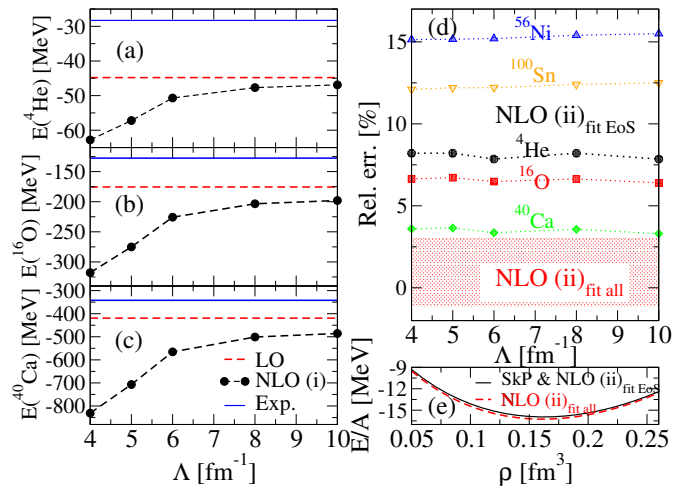}
\caption{Left panels (a)-(c): Ground state energies up to NLO for $^{4}$He, $^{16}$O and $%
^{40}$Ca as a function of $\Lambda$. 
Right panel (d): Rel. err.$=(E_b-E_{exp.})/|E_{exp.}|$ for first five closed-shell, N=Z nuclei.
The labels NLO (i) and NLO (ii)
refer to the two prescriptions described in the main text,
where parameters in NLO (ii) are obtained by either fitted to EoS only (fit EoS) or an overall fit including EoS and all five nuclei (fit all).
 Their corresponding EoS are plotted in (e).}
\label{figNLO}
\end{figure}

Once $k_{A}$ is known, we have all the ingredients to perform actual
calculations. In Refs. \cite{yang2017-b,Burrello2020}, the full $t_{0}-t_{3}$
LO interaction is iterated to generate $E_{iter}^{NLO}$ for nuclear matter. 
The same procedure can be performed in principle in Eq. (\ref{nlo3}) for finite
nuclei. 
However, some conceptual subtleties arise regarding how to account for the density $\rho$ 
when one considers the fluctuation of the wavefunctions due to the intermediate excitations. In fact, in conventional EDF
approaches with a density-dependent term included (for example the $t_3$ term of Skyrme interactions), the 
interaction does not correspond to a genuine Hamiltonian. The iteration of this term generates a conceptual drawback and may lead to technical problems such as 
divergences in BMF calculations for nuclei \cite{l1,l2,l3}.
Also, the density-dependent term depends on the wavefunction and this could potentially complicate an EFT
analysis. 
Therefore, we choose not to
iterate the $t_{3}$ part of the interaction in this work. 

In the following, we perform
two types of NLO calculations:
\begin{enumerate}[label=(\roman*)]
\item Only the $t_{0}$ part of the LO interaction
is iterated, and $V^{CT}=C(1+x_{c}P_{\sigma })$. 
\item Same
as (i), but with additional $V_{ii}^{CT}=\frac{1}{2}t_{1}(1+x_{1}P_{\sigma })(%
\mathbf{k}^{\prime 2}+\mathbf{k}^{2})+t_{2}(1+x_{2}P_{\sigma })\mathbf{k}%
^{\prime }\cdot \mathbf{k}$, that is, the Skyrme-type $t_{1,2}$ terms are added.
%\item Same as (ii), but $V^{CT}=C(1+x_{c}P_{\sigma })\rho ^{2\alpha }$.
\end{enumerate}
Note that the above interactions are Skyrme-like, and $P_{\sigma }=(1+{\sigma }_{1}{\sigma }_{2})/2$ is the
spin--exchange operator. We treat $C$, $x_c$, $\alpha$, $t_{0,1,2,3}$, and $x_{0,1,2,3}$ as the low-energy constants (LECs) in EFT, and we choose to renormalize them to reproduce
the SLy5 SM and neutron matter (NM) EoSs. The LECs, the $\chi^2$ values and 
the resulting EoSs are given in the supplemental materials.
Predictions on g.s. energies of $^{4}$He, $^{16}$O, $^{40}$Ca, $^{56}$Ni and $^{100}$Sn evaluated up to NLO
with $\Lambda =4-10$ fm$^{-1}$, are given in Fig.\ref%
{figNLO}, where the empirical 
$\hbar \protect\omega=45A^{-1/3}-25A^{-2/3} $ are adopted. As one can see, the pathological overbinding trend at LO seems to persist under the prescription (i).
%In fact, already in the uniform case, it is very difficult to obtain good simultaneous fits for both SM and NM.
Thus, without the entrance of new $k_F$-dependencies in the EoS (other than terms behave asymptotically $\sim k_F^4$ and proportional to $t_0^2$) at NLO, one does not observe any improvement from LO to NLO for both the EoS of matter and finite nuclei.
%Since one expects results up to NLO to be better than LO in a consistent EFT,
%something must be missing under prescription (i).
Nevertheless, the NLO renormalizability is satisfied---which is reflected in the converging pattern of NLO (i) results against $\Lambda$.
A real improvement is achieved by the prescription (ii), where, by just fitting to the empirical EoSs, reasonable reproductions
of the experimental binding $E_{exp.}$ are found for nuclei up to mass number $A=40$ (with $|$relative error$|=\frac{|E_b-E_{exp.}|}{|E_{exp.}|}\leq 8\%$, where $E_b$ is the resulting binding energy). 
This suggests that the $t_1$, $t_2$ terms are indeed indispensable, as indicated by many phenomenological studies.
However, the error grows to $\sim15\%$ when extending the calculation to the next two $N=Z$ nuclei ($^{56}$Ni and $^{100}$Sn). Note that the curves labelled as NLO (ii)$_{\text{fit EoS}}$ are obtained by keeping the original SkP or SLy5 values of $t_{1,2,3}$, $x_{1,2,3}$ and $\alpha$, while adjusting only $t_0$, $C$ and $x_{0,c}$ to two EoSs\footnote{LECs adjusted to SkP and SLy5 EoSs produce $\leq 1\%$ difference in $E_b$ up to $^{40}$Ca, and are indistinguishable in Fig.\ref{figNLO}.}.
Up to NLO, the computational cost stays very closed to the MF calculations and is relatively small. 
Thus, we attempt a second fit (utilizing all LECs but keeping $\alpha=1/6$) to the empirical EoSs and all five nuclei. 
We found it is possible to reproduce the experimental binding for all five nuclei within $3\%$ (denoted by the red shaded area and labelled as NLO (ii)$_{\text{fit all}}$ in Fig.\ref{figNLO}), if
one allows the SM EoS to be slightly ($\leq 2 \%$) more attractive around saturation than the one produced by SkP (panel (e), Fig.\ref{figNLO}).  
%Note that, instead of fine-tuning the LECs, our focus here is to demonstrate an approach that includes subleading contributions---a non-trivial effect
%to structure of nuclei---can work as good as those best-fitted MF Skyrmes, so that we are one step further to the elimination of model-dependence.
%Finally, the purpose of prescription (iii) is to see the impact of $V^{CT}=C(1+x_{c}P_{\sigma })\rho ^{2\alpha }$, which is designed to absorb the 
%divergence if one iterates the density-dependent $t_3$ term.
%Without the entrance of once-iterated-$t_3$ term, unfavourable results are observed.

\section{Power counting: a particle-number-dependent high- and low-momentum scales}

Finally, we speculate the high- and low-momentum scale $M_{hi}$ and $M_{lo}$ in our EFT-expansion.
Since $M_{lo}$ spans from 0 to $k_F$---which varies with $A$ in a nucleus, 
a successful EFT arrangement of observables up to NLO in terms of powers series in $(M_{lo}/M_{hi})$ suggests that 
$M_{hi}$ has the following properties:
\begin{itemize}
\item It is at least larger than $k_F$, and depends on the number of particles $A$.
\item It depends on $N_{max}$ and $\hbar\omega$, at least for those nuclei where central densities are lower than the saturation density of SM. Let us denote by $A_s$ typical $A$ values for which nuclei reach the saturation density in their central region. Then $M_{hi}$ increases with $A$ for $A<A_s$.
\end{itemize}
The breakdown scale has a functional form $M_{hi}(A,\hbar\omega)$.
For $A<A_s$, the asymptotic form of the
EFT expansion is $\frac{M_{lo}}{\bar{M}_{hi}}\sim \frac{k}{\beta k_F(A)}$, where $k$ is the characteristic center-of-mass momentum scale and $\beta\gtrapprox 1$. 
On the other hand, $\beta k_F(A)\sim \bar{M}_{hi}$, that is, becomes a constant for $A > A_s$, 
where $\bar{M}_{hi}$ is a hard breakdown scale to be extracted by a Lepage-like plot \cite{harald,harald2} from NLO and next-to-next-to-leading order (NNLO) results;
$\frac{2}{3\pi^2}\bar{M}^3_{hi}$ and $\frac{1}{3\pi^2}\bar{M}^{3}_{hi}$ correspond to the highest density $\rho$ for which one can trust the EoS of SM and NM, respectively (for example, twice the saturation density of SM). 

\section{Summary}

In summary, we provide a novel framework to include BMF 
correlations order by order. With a reliable extraction of $k_{F}$,
the treatment of finite nuclei and nuclear matter can be performed on the same
footing. Investigations up to NLO are performed for five $N=Z$ closed-shell nuclei and for nuclear matter for the first time.
We have tested various arrangements of NLO corrections through renormalization-group analysis.
Note that our analysis are based 
on a trial and error procedure. Since not all possibilities are tested, our NLO prescription (ii) might still be subjected to further refinements. 
Nevertheless, the trial and error procedure carried out in present work---which checks the renormalizability of the 
in-medium loops (and therefore the self-consistency of the proposed beyond mean field corrections)---
can be repeated with different interactions in the future. Thus,
our work serves as a
starting point toward an EFT-based description of nuclei across
the entire nuclear chart. Many interesting future works including the treatment of
higher-order correlations and a full EFT power-counting analysis are in
progress.

\begin{acknowledgments}
This work was supported by the Czech Science Foundation GACR grant 19-19640S and 22-14497S, the Swedish Research
Council (Grant number 2017-04234), the European Research Council (ERC) under the European Union's Horizon 2020 research and innovation programme (Grant agreement number 758027). Computational resources were supplied by the project ``e-Infrastruktura CZ" (e-INFRA CZ LM2018140) supported by the Ministry of Education, Youth and Sports of the Czech Republic, IT4Innovations at Czech National Supercomputing Center under project number OPEN-24-21 1892, and the Swedish National Infrastructure for Computing (SNIC) at Chalmers Centre for Computational Science and Engineering (C3SE), and the National Supercomputer Centre (NSC) partially funded by the Swedish Research Council. S. B. acknowledges support from the Alexander von Humboldt foundation.
\end{acknowledgments}

\bibliography{finite_ref} 
\bibliographystyle{apsrev4-1}

\clearpage
\newpage

\section{\label{Supplementary}Supplemental Material}
\section{Two-body matrix elements at mean-field level}

In the shell model, the $j$-$j$ coupling scheme is commonly adopted. However,
the two-body Skyrme- or Gogny-type effective interactions operate in the $L$-$S$
coupling (partial-wave) scheme. Therefore, the following transformation is needed\cite%
{lawson1980theory}, 
\begin{eqnarray}  \label{jj_ls}
&&|(n_{a}l_{a}j_{a})(n_{b}l_{b}j_{b}) J J_z \rangle = \displaystyle%
\sum_{\lambda S} \displaystyle\sum_{\mu S_{z}} \gamma^{(J)}_{\lambda
S}(j_{a}l_{a};j_{b}l_{b}) \langle \lambda \mu S S_{z} | J J_z\rangle \notag \\
&&\times \ |(n_{a}l_{a}m_{a})(n_{b}l_{b}m_{b})\lambda \mu \rangle | S S_{z} \rangle ,
\end{eqnarray}
where $\langle \lambda \mu S S_{z} | J J_z\rangle$ are the standard
Clebsch-Gordan coefficients and 
\begin{eqnarray}
&&\gamma^{(J)}_{\lambda S}(j_{a}l_{a};j_{b}l_{b})= \sqrt{%
(2j_{a}+1)(2j_{b}+1)(2S+1)(2\lambda+1)} \notag \\
&&\times \ \left\{ 
\begin{array}{ccc}
l_{a} & 1/2 & j_{a} \\ 
l_{b} & 1/2 & j_{b} \\ 
\lambda & S & J \\ 
\end{array}
\right\}.
\end{eqnarray}
Here $J$ and $J_z$ denote the total angular momentum and its $z$-component, respectively, and $S$ and $S_z$ are %
 the total
intrinsic spin and its $z$-component, respectively. 

One could rewrite the two-particle
wavefuntions in the laboratory coordinates with quantum numbers ($JJ_{z}T$)
in terms of wavefuntions in relative coordinates, that is, 
\begin{eqnarray}
&\mid &(n_{a}l_{a}j_{a})(n_{b}l_{b}j_{b})JJ_{z}T\ \rangle   \notag
\label{two-body_right} \\
&=&\ \displaystyle\sum_{nlNL}\displaystyle\sum_{mM}\displaystyle\sum_{\mu
S_{z}}\displaystyle\sum_{\lambda S}\gamma _{\lambda
S}^{(J)}(j_{a}l_{a};j_{b}l_{b})\frac{1-(-1)^{S+T+l}}{\sqrt{2(1+\delta
_{n_{a}n_{b}}\delta _{l_{a}l_{b}}\delta _{j_{a}j_{b}})}}  \notag \\
&&\times \ M_{\lambda }(nlNL;n_{a}l_{a}n_{b}l_{b})\ \langle lmLM|\lambda \mu
\rangle \ \langle \lambda \mu SS_{z}|JJ_{z}\rangle \quad   \notag \\
&&\times \ |nlm\rangle \ |NLM\rangle \ |SS_{z}\rangle \ |T\rangle ,
\end{eqnarray}
where $|T\rangle $ is the two-particle isospin eigenstate with a total
isospin $T$. Finally, the full expression of the two-body matrix element of
an effective interaction $V_{NN}$ reads 
\begin{eqnarray}
&&\langle (n_{a}l_{a}j_{a})(n_{b}l_{b}j_{b})JJ_{z}T\ |\ V_{NN}\
|(n_{c}l_{c}j_{c})(n_{d}l_{d}j_{d})JJ_{z}T\ \rangle   \notag
\label{two-body} \\
&=&\ \displaystyle\sum_{n^{\prime }l^{\prime }N^{\prime }L^{\prime }}%
\displaystyle\sum_{nlNL}\displaystyle\sum_{m^{\prime }M^{\prime }}%
\displaystyle\sum_{mM}\displaystyle\sum_{\mu ^{\prime }S_{z}^{\prime }}%
\displaystyle\sum_{\mu S_{z}}\displaystyle\sum_{\lambda ^{\prime }S^{\prime
}}\displaystyle\sum_{\lambda S}\tilde{\gamma}_{\lambda ^{\prime }S^{\prime
}}^{(J)}(j_{a}l_{a};j_{b}l_{b}) \notag \\
&&\times \  \tilde{\gamma}_{\lambda
S}^{(J)}(j_{c}l_{c};j_{d}l_{d})  
 M_{\lambda ^{\prime }}(n^{\prime }l^{\prime }N^{\prime }L^{\prime
};n_{a}l_{a}n_{b}l_{b}) \notag \\
&&\times \   M_{\lambda }(nlNL;n_{c}l_{c}n_{d}l_{d})  
\langle l^{\prime }m^{\prime }L^{\prime }M^{\prime }|\lambda
^{\prime }\mu ^{\prime }\rangle \notag \\ 
&&\times \  \langle \lambda ^{\prime }\mu ^{\prime
}S^{\prime }S_{z}^{\prime }|JJ_{z}\rangle \langle lmLM|\lambda \mu \rangle \
\langle \lambda \mu SS_{z}|JJ_{z}\rangle \quad   \notag \\
&&\times \ \langle T|\ \langle S^{\prime }S_{z}^{\prime }|\ \langle
N^{\prime }L^{\prime }M^{\prime }|\ \langle n^{\prime }l^{\prime }m^{\prime
}|\ \ V_{NN}\ |nlm\rangle \ |NLM\rangle \notag \\
&&\times \   |SS_{z}\rangle \ |T\rangle ,
\label{eq4}
\end{eqnarray}
with $\tilde{\gamma}_{\lambda S}^{(J)}(j_{a}l_{a};j_{b}l_{b})=\frac{%
1-(-1)^{S+T+l}}{\sqrt{2(1+\delta _{n_{a}n_{b}}\delta _{l_{a}l_{b}}\delta
_{j_{a}j_{b}})}}\gamma _{\lambda S}^{(J)}(j_{a}l_{a};j_{b}l_{b})$, where
symbols with prime refer to the left vector. Note that if $V_{NN}$ consists only of $s$-waves, then the kernel 
$\langle n^{\prime }l^{\prime }m^{\prime }\;|\;V_{NN}\;|\;nlm\rangle$ can be further simplified into:
%\begin{eqnarray}
%\langle n^{\prime }l^{\prime }m^{\prime }\;|\;V_{NN,12}\;|\;nlm\rangle 
%=\frac{1}{(2\pi)^6}\int_{0}^{\infty }d^3 \mathbf{p}\int_{0}^{\infty }d\mathbf{p^{\prime }}\psi _{n^{\prime }l^{\prime }m^{\prime }}(\mathbf{p})V_{NN,12}(\mathbf{p},\mathbf{p^{\prime }})\psi _{nlm}(\mathbf{p^{\prime }}), 
%\end{eqnarray}%
\begin{eqnarray}
&&\langle n^{\prime }0m^{\prime }\;|\;V_{NN,12}\;|\;n0m\rangle \notag \\ 
&=&\frac{1}{%
2\pi ^{2}}\int_{0}^{\infty }p^{2}dp\int_{0}^{\infty }p^{\prime
}{}^{2}dp^{\prime }\psi _{n^{\prime }0m^{\prime }}(p)V_{0,
0s=j}(p,p^{\prime })\psi _{n0m}(p^{\prime }). \notag \\
\end{eqnarray}
Here $V_{0,0s=j}(p,p^{\prime })$ is the $s$-wave component of $V_{NN}$, and $\psi _{nl=0m}$ is the standard Harmonic oscillator (HO) wavefuntion with
quantum numbers $n,l,m$. In the above expression, we have 
assumed that $V_{NN}$ is density independent. 
For density-dependent effective
interactions (such as the Skyrme one), one may follow Eqs. (4)-(7) in
Ref.\cite{Jiang2018} to perform the integral over the CM coordinate space $%
\mathbf{R}=(\mathbf{r_{1}}+\mathbf{r_{2}})/2$ and to evaluate the matrix element
of $\langle N^{\prime }L^{\prime }M^{\prime }|\hat{\rho}(\mathbf{R}%
)|NLM\rangle $. Here $\hat{\rho}(\mathbf{R})$ is proportional to the modular
square of the CM part of the wavefuntion, with details given in Ref.\cite%
{Jiang2018}.

The LO interaction adopted in this work has the following
form: 
\begin{equation}
V_{12}^{LO}=\delta (\mathbf{r_{1}}-\mathbf{r_{2}}%
)t_{0}(1+x_{0}P_{\sigma })+\delta (\mathbf{r_{1}}-\mathbf{%
r_{2}})\frac{t_{3}}{6}(1+x_{3}P_{\sigma })\rho ^{\alpha },  \label{vlo}
\end{equation}%
where $t_{0}$, $t_{3}$, $x_{0}$, $x_{3}$, and $\alpha $ are parameters, $%
\rho =\rho (\frac{\mathbf{r_{1}}+\mathbf{r_{2}}}{2})$ is the
density, and $P_{\sigma }=(1+{\sigma }_{1}{\sigma }_{2})/2$ is the
spin--exchange operator.

Note that Eq. (\ref{eq4}) can be reduced to a
simpler form in the case of infinity nuclear matter, where the system becomes homogeneous so that it can be described
by a Fermi gas with wavefunction $\Psi $ consisting of free wave-packets $u_{k}
$ labelled by the momentum $k$. Within a box volume $\Omega $,
Eq. (\ref{eq4}) becomes 
\begin{eqnarray}
\sum_{u_{k},u_{p}}\sum_{JT}(2T+1)(2J+1)\left\langle
u_{k}|V_{NN}|u_{p}\right\rangle ,
\end{eqnarray}%
where the sum up to the highest occupied orbitals can be converted into
integrals up to the Fermi momentum $k_{F}$, that is  
\begin{widetext}
\begin{eqnarray}
\sum_{u_{k},u_{p}}\!\!\left\langle u_{k}|V_{NN}|u_{p}\right\rangle =\frac{%
\Omega ^{2}}{(2\pi )^{6}}\int_{0}^{k_{F}}\!\!\!\!\!\!k^2dk\int d\theta_k\int d\phi_k%
\int_{0}^{k_{F}}\!\!\!\!\!\!p^2dp\int d\theta_p\int d\phi_p\frac{V_{NN}(\mathbf{k},\mathbf{p}%
)}{\Omega }.\label{eosmf} 
\end{eqnarray}
\end{widetext}

\section{Once-iterated matrix elements}

The matrix elements of the interaction given in the main text have exactly the same form as
Eq. (\ref{eq4}) here, by replacing the inner
kernel $\langle N^{\prime}L^{\prime}M^{\prime}| \ \langle
n^{\prime}l^{\prime}m^{\prime}| V_{NN} | nlm\rangle \ |N L M\rangle$ with  
\begin{eqnarray}
&&\frac{1}{4}\sum_{N^{\prime\prime}L^{\prime\prime}M^{\prime\prime}n^{\prime%
\prime}l^{\prime\prime}m^{\prime\prime}}\langle
N^{\prime}L^{\prime}M^{\prime}| \langle n^{\prime}l^{\prime}m^{\prime}|
V_{iter}^{LO} | n^{\prime\prime}l^{\prime\prime}m^{\prime\prime}\rangle
|N^{\prime\prime}L^{\prime\prime}M^{\prime\prime}\rangle  \notag \\
&&\times \ G\langle
N^{\prime\prime}L^{\prime\prime}M^{\prime\prime}| \langle
n^{\prime\prime}l^{\prime\prime}m^{\prime\prime}| V_{iter}^{LO} | nlm\rangle
|N L M\rangle,  \label{it}
\end{eqnarray}
where $G$ is given by Eq. (6) in the main text. 
For a density-independent interaction, the non-vanishing matrix elements in Eq. (\ref{it})
have quantum numbers $N^{\prime}=N$, $L^{\prime}=L$ and $M^{\prime}=M$. Furthermore, for
the $s$-wave part of the effective interaction $V_{iter}^{LO}=t_{0}(1+x_{0}P_{\sigma
})$ considered in this work, we have $n^{\prime}=n$ and $l^{\prime}=l=m^{%
\prime}=m=0$ due to the energy conservation property of the Moshinsky
transformations. In addition, one can further drop the summation regarding the CM
intermediate states $\langle
N^{\prime\prime}L^{\prime\prime}M^{\prime\prime}|$ as their overlap is
always 1. Thus, the inner kernel (excluding the external $\langle N L M|$
and $|N L M\rangle$) becomes: 
\begin{eqnarray}
&&\frac{1}{4}\sum_{n^{\prime\prime}00} \langle n00| V_{iter}^{LO} |
n^{\prime\prime}00\rangle G \langle n^{\prime\prime}00| V_{iter}^{LO} |
n00\rangle \notag \\
&=&\frac{1}{4}\sum_{\alpha} \langle n00| V_{iter}^{LO} | \alpha
\rangle G \langle \alpha| V_{iter}^{LO} | n00\rangle.  \label{iter1}
\end{eqnarray}%
Note that we have taken the freedom to re-express the intermediate states
from the HO-basis in relative coordinates into any complete set $%
\sum_{\alpha} | \alpha \rangle \langle \alpha| $ within the same coordinates,
where $| \alpha \rangle$ labels the eigenstates. One can
then choose the new basis to be the kinetic eigenstates  so that $| \alpha \rangle=| u_{k^{\prime}} \rangle
$ and the summation $\sum_{\alpha}$ is converted into an integral over the
intermediate momentum $k^{\prime}$. At the same time, one can also decompose
each outer HO-wavefuntion $\langle n00|k\rangle$ into a linear combination
of $\sum_{k}c_k u_{k}$, with $c_k=\psi^{HO}_{n00}(k)$. In this way, Eq. (\ref%
{iter1}) becomes 
\begin{eqnarray}
-\frac{m}{4}\frac{(4\pi)^2}{(2\pi)^6}\int_0^{k_F} k^2 dk \int_0^{\Lambda}
k'^2 dk^{\prime} 
\frac{\psi^{HO}_{n00}(k) [t_{0}(1+x_{0}P_{\sigma
})]^2 \psi^{HO}_{n00}(k)}{k'^2-k^2}. \notag \\ \label{iter2}
\end{eqnarray}%
Note that, together with the outer integral on $\langle N L M | $, Eq. (\ref%
{it}) becomes
\begin{widetext} 
\begin{eqnarray}
-\frac{m}{4}\frac{(4\pi)^3}{(2\pi)^9}\left[\int_0^{2k_F} K^2 dK \int_0^{k_F}
k^2 dk \int_0^{\Lambda} k'^2 dk^{\prime}\frac{\Psi^{HO}_{NLM}(K)%
\psi^{HO}_{n00}(k) [t_{0}(1+x_{0}P_{\sigma })]^2
\psi^{HO}_{n00}(k)\Psi^{HO}_{NLM}(K)}{k'^2-k^2}\right] _{BC}.
\label{iter3}
\end{eqnarray}
\end{widetext}
The $dK$ integral is not equal to 1 and the denominator does not diverge,
 due to the fact that the three variables $\mathbf{k}=(\mathbf{k_1}-%
\mathbf{k_2})/2$, $\mathbf{k^{\prime}}=(\mathbf{k^{\prime}_1}-\mathbf{%
k^{\prime}_2})/2=(\mathbf{k_1}-\mathbf{k_2})/2+\mathbf{q}$, and $\mathbf{K}=%
\mathbf{k_1}+\mathbf{k_2}=\mathbf{k^{\prime}_1}+\mathbf{k^{\prime}_2}$ are
restricted by the following BC: 
\begin{eqnarray}
|\mathbf{k_1}|<k_F, |\mathbf{k_2}|<k_F, \\
|\mathbf{q}+\mathbf{k_1}|>k_F, |\mathbf{k_2}-\mathbf{q}|>k_F.  \notag
\label{bc}
\end{eqnarray}
A detailed illustration of the above BC and the related treatments to
perform the triple integral can be found in Fig.2 and Fig.3 of Ref.\cite%
{baranger}.

We note that the results for finite nuclei obtained with the above equations
correspond to the full expressions given in Eqs. (22) and (27) in Ref.\cite%
{yang2016} for the EoS. The expressions of second-order EoSs listed in Refs.%
\cite{mog,yang2017-a,yang2017-b,Burrello2020} are the asymptotic form after
expanding the results in power series of $\Lambda $. Whereas an analytic
result of the triple integral can be obtained for the EoS of matter, Eq. (\ref{iter3}%
) can only be solved numerically for finite nuclei. Although the
asymptotic form agrees with the full expression at $\Lambda \rightarrow
\infty $, the discrepancies between them can be up to $10\%$ for lower cutoff values ($%
\Lambda \leq 6$ fm$^{-1})$. Therefore, we always adopt the full expression
in the nuclear matter calculations carried out throughout this work.

\section{Tables of renormalized low-energy constants}

We list below the LECs up to NLO based on the prescriptions (i)-(ii) as described in the main text. 
Note that here $C^{\Lambda}$ and $C^{\Lambda \ast }$ are related to $V^{CT}=C(1+x_{c}P_{\sigma }) $ by 
\begin{eqnarray}
C=C^{\Lambda} \notag \\
x_c=1-\frac{C^{\Lambda \ast }}{C^{\Lambda}}.
\label{cc}
\end{eqnarray}

\begin{table}[h]
\begin{tabular}{cccccc}
\hline
\hline
$\Lambda $ (fm$^{-1}$)                   & 4 & 5 & 6 & 8 & 10 \\ \hline
$t_{0}$ (fm$^{2}$)                       & 1.349 & -1.732 & -2.189 & -1.546 & -1.307  \\ 
$t_{3}$ (fm$^{2+3\alpha }$)              & 96.393 & 40.434 & 53.091 & 59.182 & 60.729  \\
$x_{0}$                                  &  -2.468  & -1.459 & -0.188 & -0.480 & -0.738  \\
$x_{3}$                                  & 4.188 & 2.050 & 0.567     & 0.479 & 0.474  \\
$\alpha$                                 & 0.0358 & 0.218 & 0.274 & 0.291 & 0.289  \\
$C^{\Lambda}$ (fm$^{2}$)                 & -10.994 & 2.578  & -0.631 & -2.616 & -2.241 \\
$C^{\Lambda \ast }$ (fm$^{2}$)           & 53.597 & 21.767 & 5.281 & 5.208 & 7.456  \\
$\chi ^{2}$                              & 15.1  & 30.2 & 289  & 334 & 356  \\
\hline
\hline
\end{tabular}%
\caption{$\chi^2$ and adjusted parameters based on the prescription (i) obtained for $\Lambda=$ values from 2 to 10 fm$^{-1}$.}
\label{fiti} 
\end{table}

\begin{table}[h]
\begin{tabular}{cccccc}
\hline
\hline
$\Lambda $ (fm$^{-1}$)                   & 4 & 5 & 6 & 8 & 10 \\ \hline
$t_{0}$ (fm$^{2}$)                       & -0.173 & -0.169 & -0.141 & -0.0868 & -0.139  \\ 
$t_{1}$ (fm$^{4}$)                       & 2.448 & 2.448 & 2.448 & 2.448 & 2.448  \\
$t_{2}$ (fm$^{4}$)                       & -2.784 & -2.784 & -2.784 & -2.784 & -2.784  \\
$t_{3}$ (fm$^{2+3\alpha }$)              & 69.747 & 69.747 & 69.747 & 69.747 & 69.747  \\
$x_{0}$                                  &  -0.243 & -0.115 & -0.312 & 1.387 & -0.051  \\
$x_{1}$                                  &  -0.328& -0.328 & -0.328 & -0.328 & -0.328  \\
$x_{2}$                                  &  -1.0& -1.0 & -1.0 & -1.0 & -1.0  \\
$x_{3}$                                  & 1.267 & 1.267 & 1.267 & 1.267 & 1.267  \\
$\alpha$                                 & 0.167 & 0.167 & 0.167 & 0.167 & 0.167  \\
$C^{\Lambda}$ (fm$^{2 }$)                & -12.40 & -12.398  & -12.427 & -12.468 & -12.410 \\
$C^{\Lambda \ast }$ (fm$^{2 }$)          & -2.556 & -2.579 & -2.573 & -2.827 & -2.605  \\
$\chi ^{2}$                              & 7.6$\cdot10^{-3}$  & 3.5$\cdot10^{-3}$ & 2.5$\cdot10^{-3}$  & 8.1$\cdot10^{-4}$ & 6.4$\cdot10^{-4}$  \\
\hline
\hline
\end{tabular}%
\caption{$\chi^2$ and parameters based on prescription (ii) obtained from $\Lambda=$ 2 to 10 fm$^{-1}$.
Note that at nuclear matter level we have adopted the empirical EoSs to be those given by the SLy5-mean-field.}
\label{fitii} 
\end{table}

\begin{table}[h]
\begin{tabular}{cccccc}
\hline
\hline
$\Lambda $ (fm$^{-1}$)                   & 4 & 5 & 6 & 8 & 10 \\ \hline
$t_{0}$ (fm$^{2}$)                       & -0.183 & -0.154 & -0.123 & -0.069 & 0.130  \\
$t_{1}$ (fm$^{4}$)                       & 1.625 & 1.625 & 1.625 & 1.625 & 1.625  \\
$t_{2}$ (fm$^{4}$)                       & -1.710 & -1.710 & -1.710 & -1.710 & -1.710  \\
$t_{3}$ (fm$^{2+3\alpha }$)              & 94.811 & 94.811 & 94.811 & 94.811 & 94.811  \\
$x_{0}$                                  &  -0.230 & -0.099 & -0.159 & -0.309 & 1.699  \\
$x_{1}$                                  &  0.653 & 0.653 & 0.653 & 0.653 & 0.653  \\
$x_{2}$                                  &  -0.537 & -0.537 & -0.537 & -0.537 & -0.537  \\
$x_{3}$                                  & 0.181 & 0.181 & 0.181 & 0.181 & 0.181  \\
$\alpha$                                 & 0.167 & 0.167 & 0.167 & 0.167 & 0.167  \\
$C^{\Lambda}$ (fm$^{2 }$)                & -14.653 & -14.678  & -14.717 & -14.781 & -14.849 \\
$C^{\Lambda \ast }$ (fm$^{2 }$)          & -10.263 & -10.327 & -10.350 & -10.410 & -10.405  \\
$\chi ^{2}$                              & 9.7$\cdot10^{-3}$  & 3.5$\cdot10^{-2}$ & 2.4$\cdot10^{-3}$  & 6.1$\cdot10^{-3}$ & 2.2$\cdot10^{-2}$  \\
\hline
\hline
\end{tabular}%
\caption{$\chi^2$ and parameters based on prescription (ii) obtained from $\Lambda=$ 2 to 10 fm$^{-1}$. 
Note that at nuclear matter level we have adopted the empirical EoSs to be those given by the SkP-mean-field.  }
\label{fitii2}
\end{table}

\begin{figure}[tbp]
\includegraphics[scale=0.4]{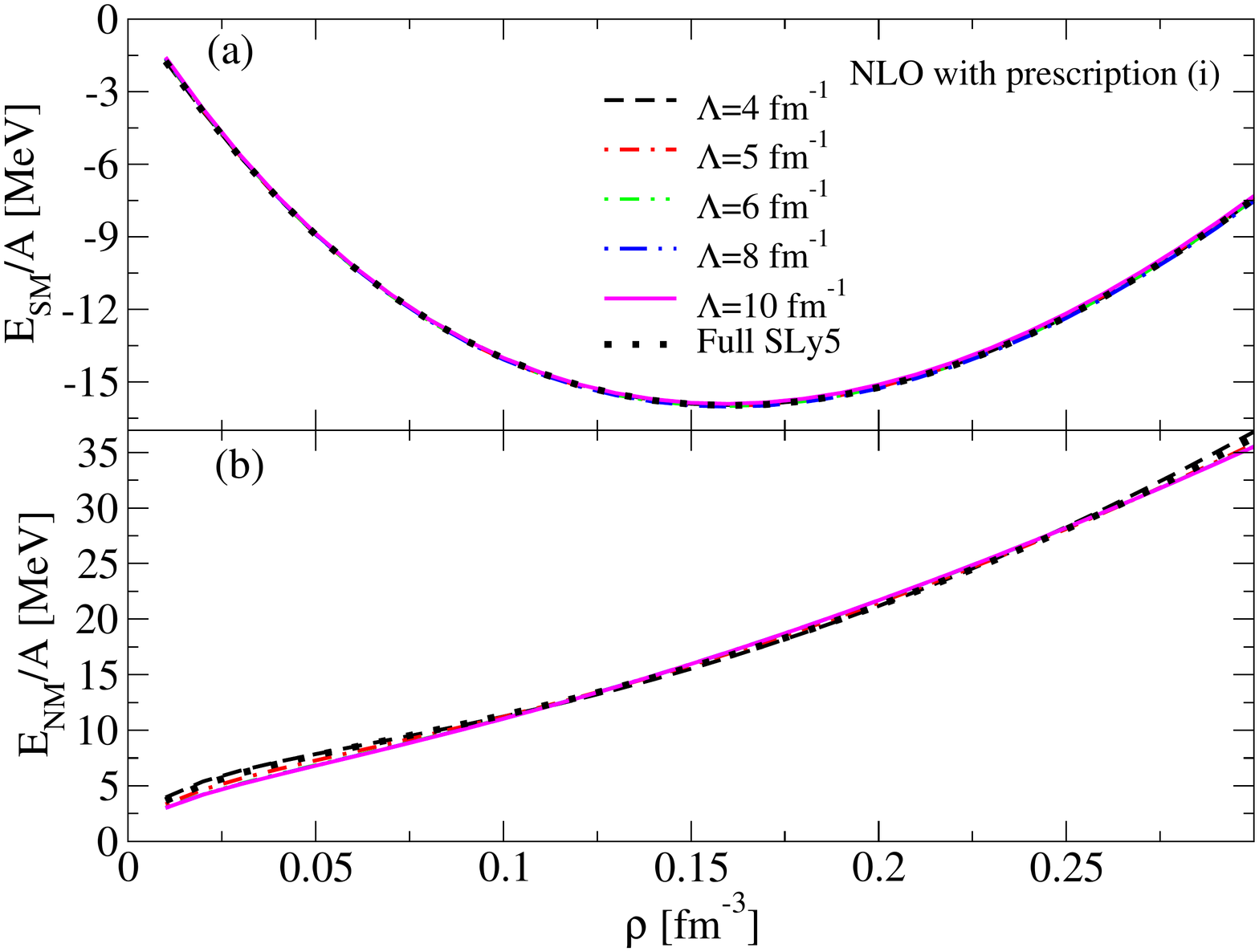}
\caption{Second–order EOSs for SM (a) and NM (b) under the prescription (i) adjusted on the full SLy5-mean-field EOSs,
with an effective mass equal to the bare mass, for different values of the cutoff. }
\label{eosnloi}
\end{figure}

\begin{figure}[tbp]
\includegraphics[scale=0.4]{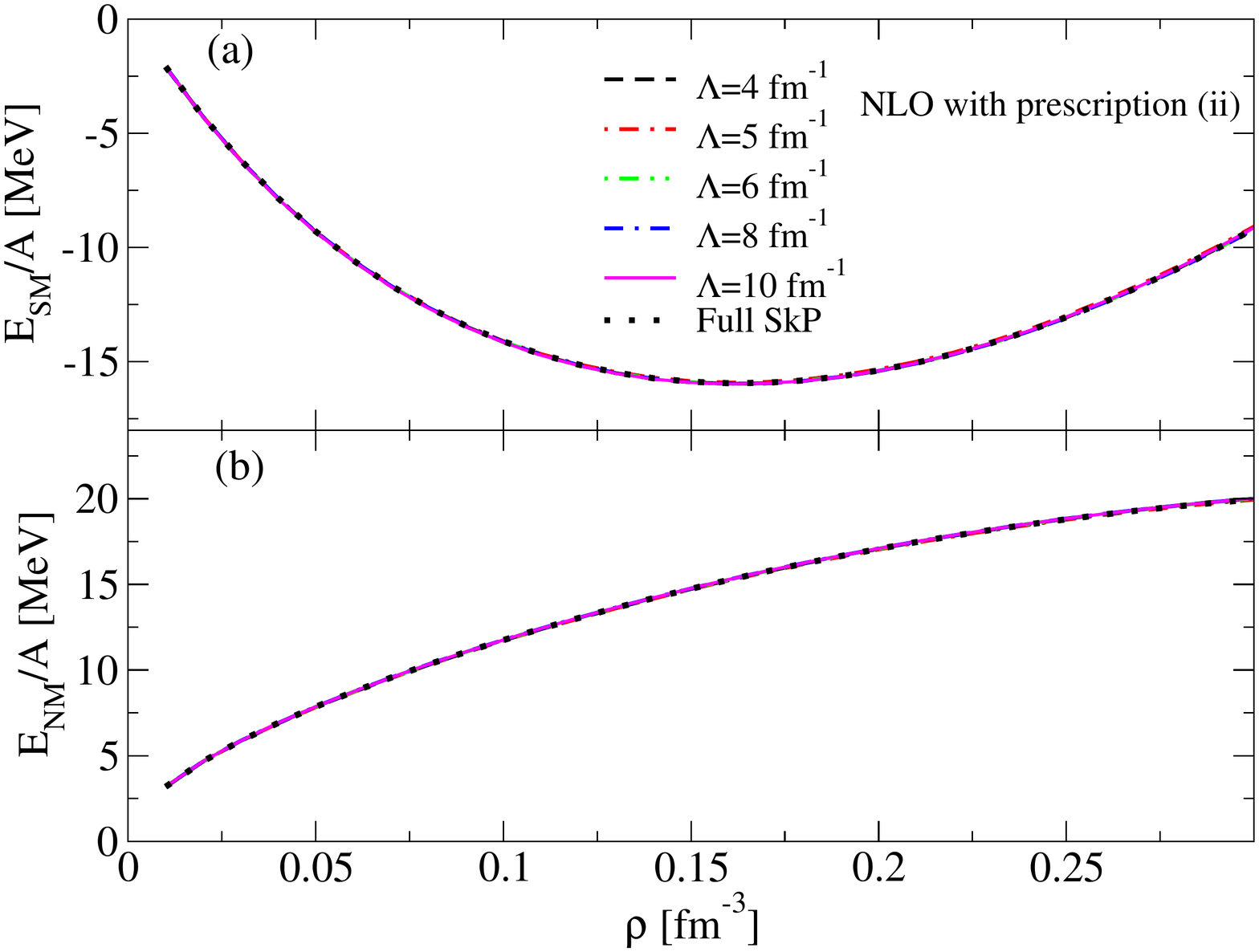}
\caption{Second–order EOSs for SM (a) and pure NM (b) under the prescription (ii) adjusted on the full SkP-mean-field EOSs,
with an effective mass equal to the bare mass, for different values of the cutoff.}
\label{eosnloii}
\end{figure}

\begin{figure}[tbp]
\includegraphics[scale=0.4]{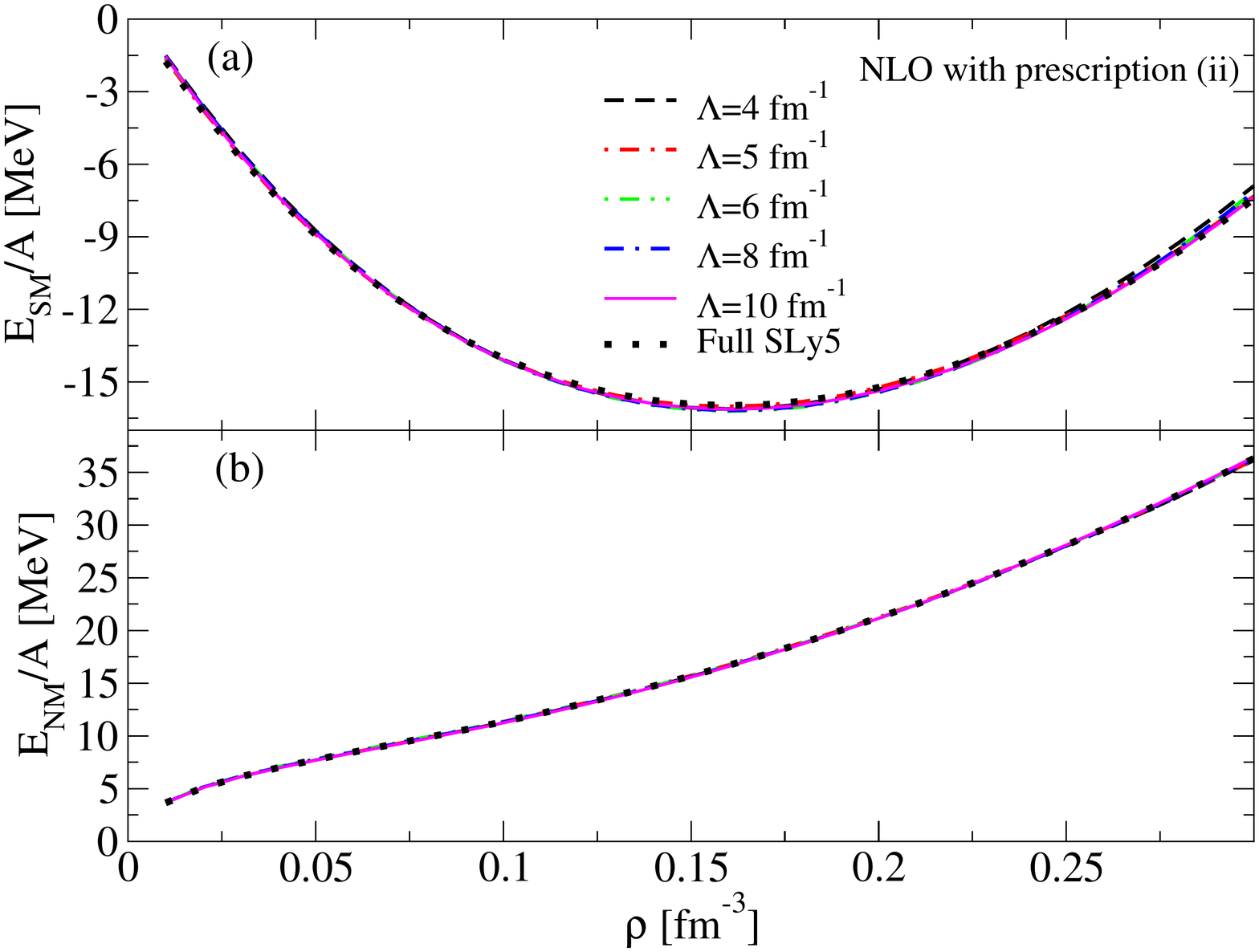}
\caption{Second–order EOSs for SM (a) and pure NM (b) under prescription (ii) adjusted on the full SLy5-mean-field EOSs,
with an effective mass equal to the bare mass, for different values of the cutoff.}
\label{eosnloii2}
\end{figure}

The corresponding EoSs generated by LECs listed in Tables \ref{fiti}-\ref{fitii2} are plotted as Figs.\ref{eosnloi}-\ref{eosnloii2}.

Finally, we list the LECs correspond to NLO (ii)$_{\text{fit all}}$ in Table IV. 
\begin{table}[h]
\begin{tabular}{ccccc}
\hline
\hline
$\Lambda $ (fm$^{-1}$)                   & 4 & 6 & 8 & 10 \\ \hline
$t_{0}$ (fm$^{2}$)                       & -4.96 & -4.36 & -3.72 & -3.62   \\ 
$t_{1}$ (fm$^{4}$)                       & -1.06 & -0.076 & 0.15 & 0.16  \\
$t_{2}$ (fm$^{4}$)                       & -7.49 & -8.79 & -9.06 & -8.61   \\
$t_{3}$ (fm$^{2+3\alpha }$)              & 35.81 & 48.09 & 45.09 & 61.96   \\
$x_{0}$                                  &  -0.20 & -0.13 & -0.62 & -0.47   \\
$x_{1}$                                  &  -3.15& -61.93 & 49.55 & 40.72   \\
$x_{2}$                                  &  -0.91& -1.02 & -1.04 & -1.04   \\
$x_{3}$                                  &  2.33 & 1.29 & 1.66 & 0.95  \\
$\alpha$                                 & 0.167 & 0.167 & 0.167 & 0.167   \\
$C^{\Lambda}$ (fm$^{2 }$)                & 16.8 & 19.1  & 27.7 & 28.1  \\
$C^{\Lambda \ast }$ (fm$^{2 }$)          & 35.2 & 33.9 & 69.2 & 65.1 \\
\hline
\hline
\end{tabular}%
\caption{Parameters of NLO (ii)$_{\text{fit all}}$ obtained from $\Lambda=$ 4 to 10 fm$^{-1}$.
Note that for all $\Lambda$, they produce symmetric EoSs which are consistently more attractive ($\sim 2\%$) than the SkP-mean-field value around saturation density, as shown in Fig.3 (e) in the main text.}
\label{fitii} 
\end{table}
\end{document}